\documentclass{emulateapj}
\usepackage{amstext,amsmath,natbib}

\slugcomment{To be published in the Astrophysical Journal, vol 612}
\shorttitle{Interstellar X-ray Spectroscopy} 
\shortauthors{JUETT ET AL.}

\begin{document}

\newcommand{\XM}{{\em XMM}}
\newcommand{\Ch}{{\em Chandra}}
\newcommand{\eps}{{\rm erg\,s^{-1}}}
\newcommand{\epcs}{{\rm erg\,cm^{-2}\,s^{-1}}}
\newcommand{\cts}{{\rm count\,s^{-1}}}

\title{High-Resolution X-ray Spectroscopy of the Interstellar Medium:
  \\ Structure at the Oxygen Absorption Edge}

\author{Adrienne~M.~Juett\altaffilmark{1}, Norbert~S.~Schulz, and
Deepto~Chakrabarty\altaffilmark{1,2}}

\affil{\footnotesize Center for Space Research, Massachusetts
Institute of Technology, Cambridge, MA 02139;\\ ajuett, nss,
deepto@space.mit.edu}

\altaffiltext{1}{Also Department of Physics, Massachusetts Institute
of Technology, Cambridge, MA 02139} 
\altaffiltext{2}{Alfred P. Sloan Research Fellow}

\begin{abstract}
We present high-resolution spectroscopy of the oxygen $K$-shell
interstellar absorption edge in seven X-ray binaries using the High
Energy Transmission Grating Spectrometer (HETGS) onboard the {\em
Chandra X-ray Observatory}.  Using the brightest sources as templates,
we found a best-fit model of two absorption edges and five Gaussian
absorption lines.  All of these features can be explained by the
recent predictions of $K$-shell absorption from neutral and ionized
atomic oxygen.  We identify the $K\alpha$ and $K\beta$ absorption
lines from neutral oxygen, as well as the $S=3/2$ absorption edge.
The expected $S=1/2$ edge is not detected in these data due to overlap
with instrumental features.  We also identify the $K\alpha$ absorption
lines from singly and doubly ionized oxygen.  The \ion{O}{1} $K\alpha$
absorption line is used as a benchmark with which to adjust the
absolute wavelength scale for theoretical predictions of the
absorption cross-sections.  We find that shifts of 30--50~m\AA\/ are
required, consistent with differences previously noticed from
comparisons of the theory with laboratory measurements.  Significant
oxygen features from dust or molecular components, as suggested in
previous studies, are not required by our HETGS spectra.  With these
spectra, we can begin to measure the large-scale properties of the
interstellar medium (ISM).  We place a limit on the velocity
dispersion of the neutral lines of $\lesssim$200~km~s$^{-1}$,
consistent with measurements at other wavelengths.  We also make the
first measurement of the oxygen ionization fractions in the ISM.  We
constrain the interstellar ratio of \ion{O}{2}/\ion{O}{1} to
$\approx$0.1 and the ratio of \ion{O}{3}/\ion{O}{1} to $\lesssim$0.1.
This work demonstrates the utility of X-ray spectroscopy for studies
of the ISM.  Future work will provide measurements of the relative
abundances and ionization fractions for elements from carbon to iron.
\end{abstract}

\vspace{0.2in}

\section{Introduction}
In the space between the stars resides the interstellar medium (ISM),
composed of gas and dust and containing both compact clouds and a
diffuse component.  The diffuse ISM has four major phases: cold
neutral, warm neutral, warm ionized, and hot ionized \citep{hk87}.
Despite years of study, questions still remain as to the relationship
between these components.  High-resolution X-ray observations from the
{\em Chandra X-ray Observatory} and {\em XMM-Newton} offer a new
window with which to study the ISM.

The ISM affects X-ray spectra in two ways: photoelectric absorption,
particularly at low energies, and scattering by dust grains, producing
X-ray halos.  At X-ray energies, absorption features are from the
excitation and ionization of inner-shell ($K$-shell) electrons,
although for high-$Z$ elements like iron, $L$-shell absorption edges
are also detectable.  The wavelength range available to \Ch\/ and
\XM\/ includes absorption features from carbon through iron.  Since the
abundances of these elements are substantially lower than for hydrogen
and helium, X-ray studies can probe the ISM over larger distances than
is possible in the optical and ultraviolet.  Measurements of the ISM
absorption features in the spectra of X-ray binaries will allow us to
study the abundances and ionization fractions for a large number of
elements.  This study is analogous to optical and ultraviolet
observations of stars and quasars that measure ISM and intergalactic
medium absorption features.  X-ray binaries are the brightest objects
in the X-ray sky and therefore require only moderate observation
lengths to produce high signal-to-noise spectra.

An understanding of the ISM absorption features is also important for
determining the intrinsic spectral properties of X-ray sources.  Early
photoelectric absorption models assumed that the ISM was composed of a
neutral, monoatomic, homogeneous gas with solar abundances \citep[see
e.g.,][]{bg70}.  \citet{rw77} were the first to compute a more complex
2-phase photoelectric absorption model which more closely resembles
the actual ISM.  Improvements in X-ray absorption cross-sections, and
solar and ISM abundance data, led to updated absorption models
\citep*{mm83,bm92,wam00}.

While the recent models provide good fits to low- and
medium-resolution X-ray spectra, they are not appropriate for the
grating spectra available with {\em Chandra} and {\em XMM}.  Most
importantly, the absorption cross-sections utilized in the standard
models \citep*{hgd93,vyb+93,vy95} do not accurately describe the
resonance structure of the edges.  Instead, the cross-sections are
modeled by a step function at a fixed edge position.  The positions
were determined either from low-resolution solid-state data
\citep{hgd93} or from theoretical calculations \citep{vyb+93,vy95},
neither of which can be reliably taken as the true atomic edge
position.  Additionally, the models do not include any contribution
from ionized forms of the elements.  The limitations of the standard
absorption models are well known \citep[see][]{wam00}, yet no
high-resolution model is available for standard usage.

The first attempt to measure ISM absorption edges in the X-rays used
the {\em Einstein} Focal Plane Crystal Spectrometer \citep{sc86} and
found evidence for the \ion{O}{1} $1s$-$2p$ ($K\alpha$) absorption
resonance and a possible \ion{O}{2} edge.  Recent \Ch\/ and \XM\/
results have revealed more complex structure around the oxygen
$K$-shell absorption edge.  Three studies have used \Ch\/ to study ISM
absorption edges in X-ray binary spectra \citep{pbv+01,scc+02,tfm+02}.
Each found a prominent \ion{O}{1} $1s$-$2p$ resonance absorption line
at 23.51~\AA\/ and evidence for another line feature at
$\approx$23.36~\AA, attributed to compound forms of oxygen, but also
consistent with the \ion{O}{2} $1s$-$2p$ transition. \citet{tfm+02}
found that the oxygen edge in Cyg X-2 could be best fit by a model
consisting of three absorption edges, representing the two edges in
atomic oxygen absorption and a contribution from oxygen in molecular
forms.  Using the {\em XMM} Reflection Grating Spectrometer,
\citet{ddk+03} found that the ISM oxygen $K$-shell edge of both X-ray
binaries and extragalactic sources was well described by the
theoretical R-matrix calculation of atomic oxygen \citep{mk98},
although requiring a shift of the edge complex relative to the
$1s$-$2p$ transition.  All of the high-resolution work points to a
complex system of absorption features at the oxygen edge, not the
simple step function used in absorption models.

The purpose of this work is to compile a large sample of spectra to
study the oxygen $K$-shell edge complex.  Other ISM edges will be
studied in a future work.  One of our goals is to provide the
community with a reasonable model for the oxygen $K$-shell edge
complex for use in future analyses.  The high-resolution astrophysical
data can also be used to test our understanding of the atomic
structure that leads to features in photoelectric absorption by
comparing with theoretical calculations.  Finally, we will study the
nature of the ISM by measuring the abundances in various ionizations
states and studying the uniformity of the ISM absorption.  Combined
with similar studies of other edges, high-resolution X-ray spectra
will in the future probe relative abundances and depletion in the ISM.

\section{Expected Structure of the Oxygen Absorption Edge}
The interstellar absorption features found in the X-ray band are due
to ionization of inner-shell ($n=1$) electrons in neutral atoms.  This
absorption results in a series of absorption edge complexes, one for
each element.  The cross-section for each edge has a similar
large-scale energy dependence.  Photons with energies below the
ionization potential cannot ionize the atom, and the cross-section is
negligible.  At the ionization energy, the cross-section is large and
then decreases roughly as $E^{-2.7}$ at higher energies.  The
reduction at higher energies is due to the increasing probability that
a photon has enough energy that it just passes by the atom without
ionizing it.  This description is employed in the standard absorption
models and is reasonable for CCD-resolution spectra.

\begin{figure}
\epsscale{1.2} \plotone{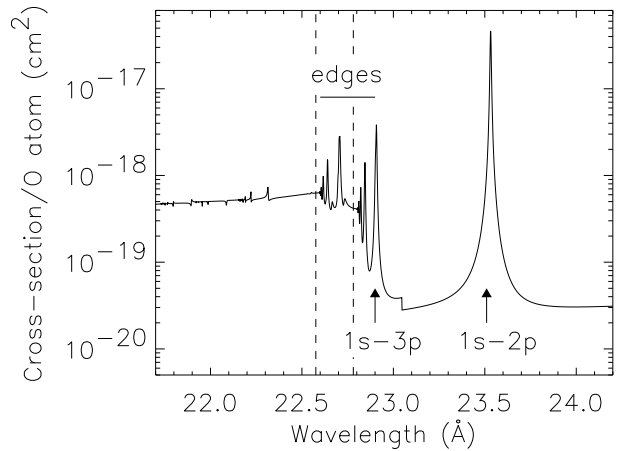} \figcaption{Calculated theoretical
cross-section of neutral oxygen, from \citet{gm00}.  Labeled are the
positions of the $1s$-$2p$ and $1s$-$3p$ transitions and the edges.
The dotted lines show the positions of the $S=3/2$ and $S=1/2$
resonance series limits at 22.781 and 22.576~\AA, respectively.}
\label{fig:1}
\end{figure}

But with the higher-resolution spectra available from grating
instruments, we can resolve finer structure.  The most important
effect is due to excitation of a $K$-shell electron into a higher
orbital.  Photons with energy slightly less (or wavelengths greater)
than the ionization energy can excite the electron into a higher $n$
shell.  This produces narrow absorption lines at the excitation
energies yielding the $1s$-$2p$ ($K\alpha$), $1s$-$3p$ ($K\beta$),
etc., features.  The $1s$-$np$ series of lines is termed a resonance
series, ending at the series limit or ionization energy
($n\rightarrow\infty$).  More structure in the edge results from
considering the spin of the final state of the atom.  For oxygen, the
ionized O$^{+}$ product with a $1s$ hole can have total spin angular
momentum $S=1/2$ or $S=3/2$, producing separate series limits at
different energies due to the spin-dependent exchange force.  Each
series limit has an associated resonance series from neutral oxygen
excitations which have the same configuration of the inner shells
($n=1,2$ with a $1s$ hole), neglecting the contribution of the valence
($n\geq3$) electron.  In addition, differences in the total orbital
angular momentum $L$ of the final state also give rise to different
resonance series.  The dominant final states have $L=1$ with $S=1/2$
and $S=3/2$, the configurations $1s 2s^2 2p^4$($^{2}P$) and $1s 2s^2
2p^4$($^{4}P$), respectively.  There are also additional series
associated with the $L=0$ and $L=2$ final states,
$1s2s^22p^4$($^{2}S$) and $1s2s^22p^4$($^{2}D$), respectively
\citep[see][]{mk98}, but these will produce weaker features since they
require both a promotion of a $1s$-$np$ electron and a transition in
the $2p^4$ subshell.  These weaker features are undetectable with
current X-ray observatories.

To calculate the theoretical photoelectric absorption cross-sections,
one must solve the Schr\"{o}dinger equation.  For atoms with only one
electron (hydrogen-like) this calculation is straightforward and
exactly solvable, but as the number of electrons in the atom
increases, so does the complexity of the calculation.  Multi-electron
atoms require numerical methods to solve for the cross-sections.  Such
methods are only able to approximate the solution of the
Schr\"{o}dinger equation.  There are a number of numerical techniques
employed \citep[see][for a review]{b00}.  The preferred method for
calculating cross-sections that include resonance structure is the
R-matrix method.  While the structure and strength of the
cross-sections calculated using the R-matrix method compare well with
laboratory measurements, the absolute positions are different.  This
results from the truncation of the wavefunction description of the
resulting ion, which is necessary to keep the computation manageable.
In practice, the energies of the theoretical cross-sections are
shifted when compared to laboratory data.

\begin{deluxetable}{lcc}
\tablecaption{Theoretical Predictions for Oxygen $K$ Edge Features}
\tablehead{ & \colhead{Predicted} & \colhead{Shifted} \\ \colhead{Feature} & 
  \colhead{$\lambda$ (\AA)} & \colhead{$\lambda$ (\AA)}
}
\startdata
\multicolumn{3}{c}{Gorczyca \& McLaughlin 2000} \\ \tableline
$1s$-$2p$ & 23.532 & 23.508\tablenotemark{a} \\
$1s$-$3p$ & 22.907 & 22.884 \\
$1s2s^22p^4$ ($^4P$) & 22.781 & 22.758 \\
$1s2s^22p^4$ ($^2P$) & 22.576 & 22.553 \\ \tableline
\multicolumn{3}{c}{Pradhan et al. 2003} \\ \tableline
\ion{O}{1} $1s$-$2p$ & 23.45 & 23.508\tablenotemark{a} \\
\ion{O}{2} $1s$-$2p$ & 23.27 & 23.33 \\
\ion{O}{3} $1s$-$2p$ & 23.08 & 23.14 \\
 & 23.02 & 23.08 \\
 & 22.93 & 22.99 \\
\ion{O}{4} $1s$-$2p$ & 22.73 & 22.79 \\
 & 22.67 & 22.73 \\
 & 22.46 & 22.52 \\
\vspace{-0.1in}
\enddata
\tablenotetext{a}{Referenced to weighted mean of observational
results, see \S~5.}
\end{deluxetable}

Motivated by the high-resolution spectroscopy possible with \Ch\/ and
\XM, several groups have published theoretical calculations of the
cross-sections of inner-shell processes in atomic and ionized oxygen
\citep{mk98,gm00,pcd+03}.  As mentioned above, the cross-section of
atomic oxygen in the vicinity of the $K$ edge is dominated by two
resonance series converging to the limits corresponding to the $^2P$
and $^4P$ states of O$^{+}$ \citep[see e.g.,][]{mk98}.  Laboratory
data \citep{mbk+96,sls+97,mk98} compare well qualitatively to these
calculations, but discrepancies in the positions of the resonance
features were noted by both \citet{sls+97} and \citet{mk98}.  More
recent work \citep{gm00} has improved on the theoretical calculations
in order to better match the data.  \citet{gm00} adjusted their
theoretical calculation to match the threshold energies found by
\citet{sls+97}, producing theoretical data that matches laboratory
measurements well in both cross-section and position of the resonance
features.  It should be noted that while laboratory measurements have
very good relative accuracy, the absolute positional accuracy is less
well known.  We therefore find it reasonable to suggest that the
energy scale of the laboratory measurements, and consequently the
theoretical calculations, may require a shift to match the wavelength
calibration of the {\em Chandra} data.  Figure 1 shows the calculated
cross-section of \citet{gm00} along with identified features.  Table 1
lists the predicted wavelengths of the important resonance features,
i.e. the $1s$-$2p$ and $1s$-$3p$ resonance lines, and the two series
limits as found by \citet{gm00}, as well as the expected values of
those features when the $1s$-$2p$ transition is shifted to the mean
value found in our analysis (see \S 5).

Along with neutral oxygen, we would expect ionized oxygen to be
present in the ISM.  \citet{pcd+03} calculated the positions and
cross-sections of the $1s$-$2p$ resonance features from all ionized
forms of oxygen.  Table 1 lists the predicted wavelengths of the
\ion{O}{1}--\ion{O}{4} lines, as well as their positions when the
\ion{O}{1} $1s$-$2p$ position is shifted to the mean value found in
our analysis (see \S 5).  The \citet{pcd+03} result shows a
significant discrepancy with the \citet{gm00} result for the predicted
wavelength of the \ion{O}{1} $1s$-$2p$ transition.  While the absolute
positions may differ due to differences in the calculation, we assume
that the relative positions of these features are reliable.  We can
test this assumption by comparing the theoretical position of the
\ion{O}{2} line to the laboratory measurement of 23.35~\AA\/
\citep{kyo+02}.  The position given by \citet{pcd+03} is 23.27~\AA,
but when our shift is applied, the theoretical position becomes
23.33~\AA, in much better agreement with the laboratory measurement.
We have recently become aware of new calculations of the full
$K$-shell absorption edge cross-sections for ionized oxygen (Gorczyca
2004, in prep.), but do not include these results in our analysis.

\section{Observations and the HETGS Oxygen Edge}
Our study is based on archival observations of seven bright X-ray
binaries obtained with {\em Chandra} using the High Energy
Transmission Grating Spectrometer (HETGS) in combination with the
Advanced CCD Imaging Spectrometer \citep[ACIS;][]{wbc+02}.  The HETGS
spectra are imaged by ACIS, an array of six CCD detectors.  The
HETGS/ACIS combination provides both an undispersed (zeroth order)
image and dispersed spectra from the gratings.  The various orders
overlap and are sorted using the intrinsic energy resolution of the
ACIS CCDs.  The HETGS carries two transmission gratings: the Medium
Energy Gratings (MEGs) with a range of 2.5--31~\AA\/ (0.4--5.0~keV)
and the High Energy Gratings (HEGs) with a range of 1.2--15~\AA\/
(0.8--10.0~keV).  The first-order MEG (HEG) spectrum has a spectral
resolution of $\Delta\lambda=$ 0.023~\AA\/ (0.012~\AA).

Of the previous ISM studies of the oxygen $K$-shell edge, only one
\citep{scc+02} was performed with the HETGS.  The oxygen edge is
accessible with the HETGS exclusively via the MEG.  The other studies
were performed with either the Low Energy Transmission Grating
Spectrometer (LETGS) on \Ch, or the Reflection Grating Spectrometer
(RGS) on {\em XMM}.  The MEG offers the highest available spectral
resolution $\lambda/\Delta\lambda$$\approx$1000, compared with 460 for
the LETGS, and 300 for the RGS at 23~\AA, but the lowest effective
area (5, 8, and 60~cm$^{2}$ for the MEG, LETGS, and RGS respectively
at 23~\AA).  

The MEG images both the positive and negative grating orders onto
ACIS.  For the oxygen edge region of the spectrum, the $-$1 order is
readout by the back-side illuminated S1 CCD, while the $+$1 order is
readout by the front-side illuminated S4 CCD.  In the brightest
sources, we found a count excess in the $+$1 spectrum at the oxygen
edge, possibly from a previously undetected hot pixel, and have
therefore chosen to restrict our analysis to the $-$1 MEG spectrum.
Since the effective area of S1 is $\approx$6 times larger than S4,
this does not significantly affect our countrate.  The effective area
curve of the MEG $-$1 order is shown in Figure 2.  The structure in
the oxygen $K$-shell region is from internal absorption in the CCDs
and the optical blocking filter (OBF).  The feature at 23.33~\AA\/ is
likely from the $1s$-$2p$ resonance of double-bonded oxygen found in
the OBF, which is composed of polyimide (a polycarbonate plastic).  A
similar resonance feature is found in absorption spectra of O$_{2}$
(A. Hitchcock, priv. comm.).  The instrumental features were well
determined prior to the launch of \Ch\/ in 1999\footnote{For 
information on the calibration of the OBF see
http://www.astro.psu.edu/xray/docs/cal\_report/cal\_report.html .}
\citep{pbk+97}.

\begin{figure}
\epsscale{1.2} \plotone{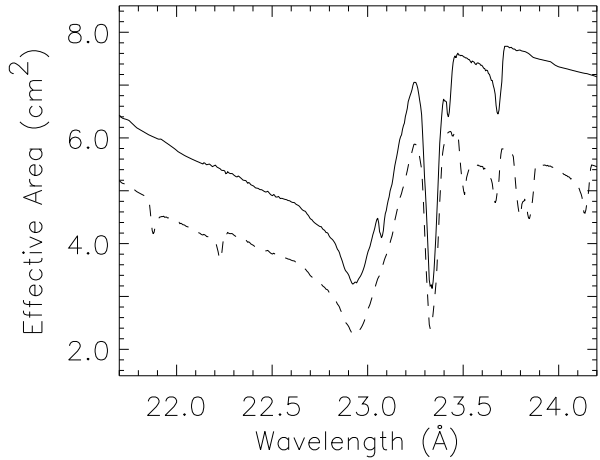} \figcaption{Effective area curves of
the MEG $-$1 order spectra on ACIS-S1 for two observations.  The solid
line is from the Cyg X-2 observation taken 1999 September 23, while
the dashed line is from the Cyg X-1 observation taken 2002 July 30
(ObsID 3724).  The difference in the effective area is due to the
contaminant which is building up on the ACIS detector.}
\label{fig:2}
\end{figure}

Along with the known instrumental contribution to the oxygen edge, a
time-dependent reduction in the effective area of ACIS has been
detected at low energies.  This reduction is due to a layer of
contaminant that is accumulating on the ACIS detector which absorbs
incoming X-rays.  The contaminant gradually reduces the detector
quantum efficiency with time and affects all wavelengths, particularly
those above 6~\AA.  The {\em Chandra} instrument teams have studied
the spectral profile of the contaminant absorption using LETGS spectra
and have modeled its structure and time-dependence \citep{mtg+03}.  An
estimated correction for the contaminant is now available through the
{\em Chandra} website\footnote{See
http://asc.harvard.edu/ciao/threads/aciscontam/}.  The biggest effect
is an overall reduction in effective area at the oxygen edge due to
the large optical depth of carbon.  Figure 2 shows the change in the
effective area over time, from 1999 September to 2002 July.  In
addition, a small ($\tau$$=$0.10 in 2002 October) oxygen edge from the
contaminant is also detected.  There is some discrepancy between the
predicted optical depth of oxygen (based on the behavior of the carbon
optical depth) and the measured oxygen optical depth, which is not
understood \citep{mtg+03}.  Since all of our observations were taken
before October 2002, the contaminant contribution to the oxygen
optical depth is $\lesssim$0.10 for all observations, which translates
to less than 10\% of the total instrumental edge.  This is comparable
to the uncertainty on our optical depth measurements (see \S 5), and
therefore our results should not be significantly affected by the
uncertainties in the contaminant contribution.  A more important issue
is the structure of the contaminant edge.  High-resolution studies
reveal structure similar to that of the OBF \citep{mtg+03}.  Thus,
besides changes in the instrumental optical depth, the contaminant
also adds to the equivalent width of the instrumental absorption
feature at 23.33~\AA.  It should be noted that the high-resolution
structure of the oxygen edge was determined using a single observation
\citep{mtg+03}, and no estimate of the error in this measurement is
available.  We assume the contaminant correction is accurate, but see
\S 6 for possible problems with the correction.

\begin{deluxetable}{lccc}
\tablecaption{Observation Log}
\tablehead{\colhead{Source Name} & \colhead{ObsID} & 
  \colhead{Observation Date} & \colhead{Time (ks)}
}
\startdata
Cyg X-2       & 1102 & 1999-09-23 & 29 \\
Cyg X-1       & 107  & 1999-10-19 & 9  \\
              & 3407 & 2001-10-28 & 21 \\
              & 3724 & 2002-07-30 & 14 \\
4U~1636$-$53  & 105  & 1999-10-20 & 29 \\
              & 1939 & 2001-03-28 & 26 \\
4U~1735$-$44  & 704  & 2000-06-09 & 24 \\
GX 9$+$9      & 703  & 2000-08-22 & 21 \\
4U~1543$-$624 & 702  & 2000-09-12 & 27 \\
4U~1820$-$30  & 1021 & 2001-07-21 & 9  \\
              & 1022 & 2001-09-12 & 9  
\enddata 
\end{deluxetable}

In Table 2, we list the observations used in this analysis.  The data
sets were reduced using the standard CIAO
threads\footnote{http://asc.harvard.edu/ciao/threads/}.  In some
cases, the zeroth-order data was not telemetered.  For these
observations, we estimated the zeroth-order position by finding the
intersection of the grating arms and readout streak.  After spectral
extraction, the accuracy of the estimated zeroth-order position was
verified by comparing the wavelengths of strong instrumental edges in
both plus and minus sides of the spectra.  Events collected during
thermonuclear X-ray bursts were excluded from this analysis.  Since
the individual observations of 4U~1820$-$30 and 4U~1636$-$53 were
brief, the spectra from multiple observations were combined for each
source in order to improve statistics.  Detector response files (ARFs
and RMFs) were created for the $-1$ MEG spectra.  We corrected the
standard CIAO ARFs for the time-dependent change in the response due
to the contaminant on ACIS using the {\tt contamarf} tool.

\vspace{0.2in}
\section{Spectral Fitting Procedure}
We fit only the 21.5--24.5~\AA\/ wavelength range and binned the data
to ensure at least 10 counts per bin.  An absorbed power-law continuum
model was used for all sources except Cyg~X-1, where an absorbed disk
blackbody model was used.  We initially allowed the continuum
parameters to vary during the fits, but visual inspection revealed
that the best-fit parameters underestimated the continuum level at
wavelengths greater than the edge.  To better estimate the continuum
level, we fit the line-free region of the data between 23.7 and
24.7~\AA\/ and then fixed the continuum parameters to the measured
values in subsequent fits of the oxygen edge region.  We will present
results from the fixed continuum fits, as well as note differences
between fixed and free continuum methods as an estimate of the
systematic uncertainty on our results.  The {\tt tbvarabs} absorption
model \citep{wam00} was used with the oxygen abundance set to zero to
allow for an explicit modeling of the oxygen absorption.  The
equivalent hydrogen column density ($N_{\rm H}$) was fixed in all fits
to a value consistent with the measured oxygen edge depth, assuming
the \citet{wam00} oxygen abundance.

We modeled the oxygen absorption edge by using two {\tt edge} models
which mimic the structure of the theoretical determination of the edge
\citep{mk98,gm00}.  The position of the 22.6~\AA\/ edge was always
fixed.  For 4U~1735$-$44 and 4U~1543$-$624, the 22.8~\AA\/ edge
position was also fixed in the fits.  Both edge depths were always
varied, although for the lower column density sources, only upper
limits on the depth of the 22.6~\AA\/ edge were obtained.  We allowed
the two edge depths to vary independently in order to test if both
edge components are detectable as claimed by \citet{tfm+02}.  Since
the ratio of the edge depths is a fixed quantity given by the atomic
cross-section, we should find a consistent value for this ratio if
both edges are detected in our data.  This test allows us to
understand the sensitivity of our data to the edge structure.

To determine the most appropriate model for the oxygen edge region, we
used the observations of the brightest sources, Cyg X-2 and Cyg X-1
(ObsID 107) as templates (Figure 3).  Our model included absorption
lines for the 23.5~\AA\/ and 23.36~\AA\/ lines seen in other analyses
\citep{pbv+01,scc+02,tfm+02}.  For most sources the widths of these
lines were allowed to vary along with the line position and depth.  In
some cases, the widths had to be constrained to prevent
unrealistically large values.  The position of the 23.36~\AA\/ line in
the fit of the 4U~1543$-$624 spectra was fixed.  We also included two
absorption lines in the 23.0--23.2~\AA\/ wavelength range and another
at 22.8--22.9~\AA.  For these lines, we fixed the line width to the
instrumental resolution of 0.023~\AA\/ FWHM (or a sigma of 0.01~\AA),
while allowing the position and depth to find the best-fit values.
The best-fit parameters were then used to determine the equivalent
widths (EWs) of the lines.  Figures 4 and 5 show the spectra and
best-fit model for all datasets.  All error estimates are
90\%-confidence levels, unless otherwise noted, and do not include
instrumental errors.  The absolute wavelength accuracy of the MEG is
$\pm$0.011~\AA\/ FWHM.

\section{Results and Analysis}
By fitting with two edge models, we hoped to resolve the $S=1/2$ and
$S=3/2$ series limits as seen in theoretical models.  The lack of a
consistent ratio between the two edges, suggests that our data are not
good enough to consistently detect the $S=1/2$ edge.  Since this is
true, we have developed a method to determine the column density of
each edge that is independent of the model used to fit the oxygen
edge.  We used the optical depth at 21.7~\AA, calculated from our
best-fit model, and the corresponding cross-section at this wavelength
\citep[$\Delta\sigma=4.2746$$\times$$10^{-19}$~cm$^{2}$ where
$\Delta\sigma$ is the difference in cross-section from the top of the
edge to the value at $\lambda=21.7$~\AA;][]{gm00} to determine the
column density of oxygen ($N_{\rm O}$).  We estimate that this method
has a systematic error of less than 5\%.  Away from the near-edge
structure, the energy dependent cross-section of \citet{gm00} matches
the cross-section calculated by \citet{vy95}.  The \citet{hgd93}
cross-section is larger by $\approx10$\%, resulting in smaller values
of $N_{\rm O}$ and $N_{\rm H}$ when the \citet{hgd93} cross-section is
used for the same optical depth.  It is also important to note that
discrepancies in the inferred column densities may also arise when the
cross-section at the ionization wavelength is used with an edge model
placed at an arbitrary wavelength.  Using the oxygen ISM abundance of
\citet{wam00}, we calculated $N_{\rm H}$.  Table 3 gives the best fit
values for the edge positions, depths and inferred column densities in
the fixed continuum models.  In addition, we give the difference in
the calculated $N_{\rm O}$ between the fixed and free continuum
models.  These column densities do not include any correction for
oxygen bound in dust grains or other molecules.

\begin{deluxetable*}{lccccc}
\setlength{\tabcolsep}{0.1in}
\tablecaption{Oxygen Edge Parameters}
\tablehead{\colhead{Source} & \colhead{$\lambda$ (\AA)} & \colhead{$\tau$} & 
     \colhead{$N_{\rm O}$ (10$^{18}$ cm$^{-2}$)} & 
     \colhead{$\Delta N_{\rm O}$ (10$^{18}$ cm$^{-2}$)} & 
     \colhead{$N_{\rm H}$ (10$^{21}$ cm$^{-2}$)}} 
\startdata
Cyg X-2 & 22.821$\pm$0.013 & 0.55$\pm$0.11 & 1.1$\pm$0.2 & \nodata & 2.3$\pm$0.5 \\
        & 22.631 & $<$0.04 &  &  \\ \tableline
4U~1636$-$53 & 22.821$\pm$0.017 & 1.28$^{+0.3}_{-0.17}$ & 2.6$^{+1.0}_{-0.3}$ & $-0.5$ & 5.3$^{+2.1}_{-0.7}$ \\
             & 22.631 & $<$0.4 &  &  \\ \tableline
4U~1820$-$30 & 22.82$\pm$0.07 & 0.65$\pm$0.07 & 1.31$^{+0.20}_{-0.14}$ & $-0.34$ & 2.7$^{+0.4}_{-0.3}$ \\
             & 22.631 & $<$0.07 &  &  \\ \tableline
4U~1735$-$44 & 22.832 & 1.1$\pm$0.4 & 3.4$\pm$1.2 & $-1.6$ & 7$\pm$2 \\
             & 22.631 & 0.6$\pm$0.4 &  &  \\ \tableline
GX 9$+$9 & 22.86$\pm$0.02 & 1.1$\pm$0.3 & 2.2$^{+1.2}_{-0.6}$ & $-0.7$ & 4.5$^{+2.4}_{-1.2}$ \\
         & 22.631 & $<$0.5 &  &  \\  \tableline
4U~1543$-$624 & 22.832 & 1.4$\pm$0.6 & 2.8$^{+2.4}_{-1.2}$ & $+0.2$ & 6$^{+5}_{-2}$ \\
              & 22.631 & $<$1.0 &  &  \\ \tableline
Cyg X-1 & 22.854$\pm$0.017 & 2.2$\pm$0.5 & 6.3$\pm$1.4 & $-0.7$ & 13$\pm$3 \\
ObsID 107 & 22.631 & 0.9$\pm$0.5 &  &  \\ \tableline
Cyg X-1 & 22.838$\pm$0.013 & 0.70$\pm$0.10 & 2.4$\pm$0.3 & $+0.1$ & 4.9$\pm$0.6 \\
ObsID 3407 & 22.631 & 0.48$\pm$0.11 &  &  \\ \tableline
Cyg X-1 & 22.838$^{+0.017}_{-0.07}$ & 0.75$\pm$0.10 & 2.9$\pm$0.3 & $-0.3$ & 5.8$\pm$0.6 \\
ObsID 3724 & 22.631 & 0.66$\pm$0.11 &  &  
\enddata
\end{deluxetable*}

Table 4 lists the best-fit values of the line positions, widths, and
EWs of the oxygen features in the fixed continuum method, in addition
we show the difference in the EW as determined by the fixed and free
continuum methods.  The difference in EW is comparable to the
statistical error on the measurements.  Therefore, our fitting method
should not greatly affect the results.  The 23.51~\AA\/ line is
accepted to be the $1s$-$2p$ transition of neutral oxygen.  By
combining the large number of sources in our sample, we find that the
weighted mean value of the line position is 23.508$\pm$0.003~\AA.
Figure 6 shows the position determinations of the \ion{O}{1} line for
the various datasets.  Our sample includes one outlier, found in the
Cyg X-1 ObsID 107 spectrum.  An examination of the spectrum of this
source (see Figure 3) suggests that the actual line position is
higher, consistent with the other sources.  The mean value of the
\ion{O}{1} line deviates by only 0.0005~\AA\/ when this observation is
excluded from the sample.

\begin{figure}
\epsscale{1.2} \plotone{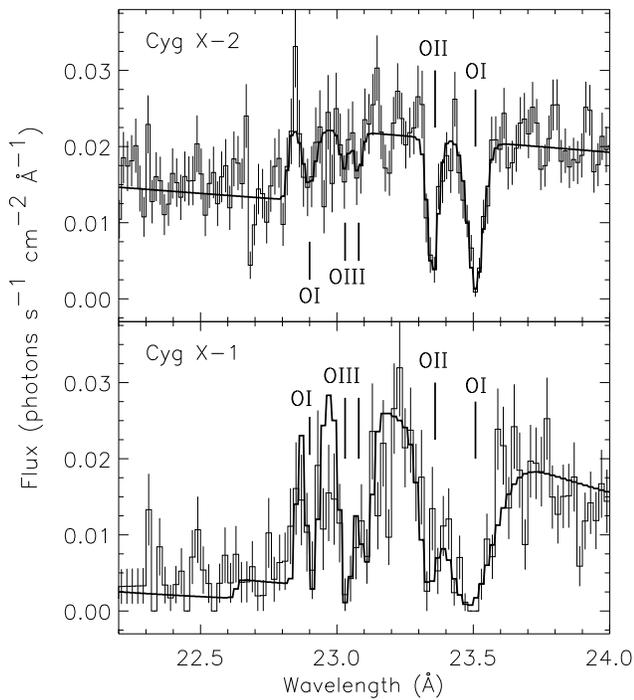} \figcaption{Flux spectra of the oxygen
edge region for Cyg X-2 and Cyg X-1 ObsID 107.  Line identifications
are also shown.}
\label{fig:3}
\end{figure}

Using our mean value for the \ion{O}{1} absorption line as a
benchmark, we can now shift all of the theoretical data to match,
allowing us to compare with the other features in the oxygen edge
region.  The weighted mean position of the first edge is
22.836$\pm$0.007~\AA.  This position is consistent with the $S=3/2$
resonance series predicted at 22.76--22.84~\AA.  In addition, the
lines with best-fit positions of 22.86--22.92~\AA\/ are reasonably
consistent with the $1s$-$3p$ resonance line of the oxygen edge at
22.884~\AA.  With the resolution of the HETG, we should resolve the
$1s$-$3p$, which must be in the spectra given the detection of the
$1s$-$2p$ line.  We therefore identify the line at 22.88~\AA\/ as the
$1s$-$3p$ resonance line from neutral oxygen absorption (see Figure 3
for identifications).

The combination of line EWs and column densities from the edges allows
us to study the curve of growth of the neutral oxygen lines.  A curve
of growth describes the relationship between the line EW and the
absorbing column density and can be separated into 3 regimes of
column density dependence: linear, flat, and square-root.  In the linear
regime, the absorption line is not saturated and the EW does not
depend on the Doppler broadening of the line, yielding a linear
dependence on the column density ($N_{\rm O}$).  As the line saturates
with increasing $N_{\rm O}$, the EW becomes explicitly dependent on
the Doppler broadening as well as on the column density.  This is
known as the ``flat part'' of the curve of growth, where the EW is
proportional to $(\ln N_{\rm O})^{1/2}$.  Finally, as the column
density increases further, the Lorentzian wings of the line profile
dominate.  In this regime, the EW is proportional to $(N_{\rm
O})^{1/2}$.  The flat part of the curve of growth is a powerful tool
for studying the Doppler broadening of the absorbing material.

Using the cross-sections of \citet{gm00}, we have calculated the
theoretical curve of growth for the $1s$-$2p$ and $1s$-$3p$
transitions of neutral oxygen with velocity dispersions of 20, 100,
and 200 km~s$^{-1}$ (Figure 7).  Overlaid are the data from this
analysis.  Both the $1s$-$2p$ and $1s$-$3p$ measured EWs are
consistent with the predicted values from the theoretical
cross-sections, which corroborates our identifications for these
lines.  Our sample of X-ray binaries probes $N_{\rm
O}$$=$$10^{18}$--$10^{19}$~cm$^{-2}$.  For the $1s$-$2p$ transition
(upper panel), this places us at the beginning of the $(N_{\rm
O})^{1/2}$ region of the curve of growth, where the contribution from
Doppler broadening is less significant.  There seems to be a
systematic offset of the data from the expected values, which is
possibly attributable to the use of a Gaussian function to fit the
lines, instead of a Lorentzian or Voigt profile.  The $1s$-$3p$
transition data points are a better match to the predicted values
(lower panel).  For this transition, we have moved fully into the flat
region of the curve of growth.  While the error in the data is
comparable to the given spread in the curve of growth due to
differences in the Doppler broadening, we can constrain the velocity
dispersion of the ISM to $\lesssim$200 km~s$^{-1}$.

For the line at $\approx$23.36~\AA, our data give a weighted mean
position of 23.351$\pm$0.003~\AA.  Previous authors have attributed
this line to the $1s$-$2p$ transition of neutral oxygen bound in
compounds \citep{pbv+01,scc+02,tfm+02}.  However, recent calculations
predict the \ion{O}{2} $1s$-$2p$ absorption line at 23.33~\AA\/
\citep{pcd+03}, and \citet{kyo+02} measure the \ion{O}{2} line at
23.35~\AA.  We therefore attribute this line to \ion{O}{2} in the ISM
(Figure 3).

\begin{figure*}
\epsscale{1.02} \plotone{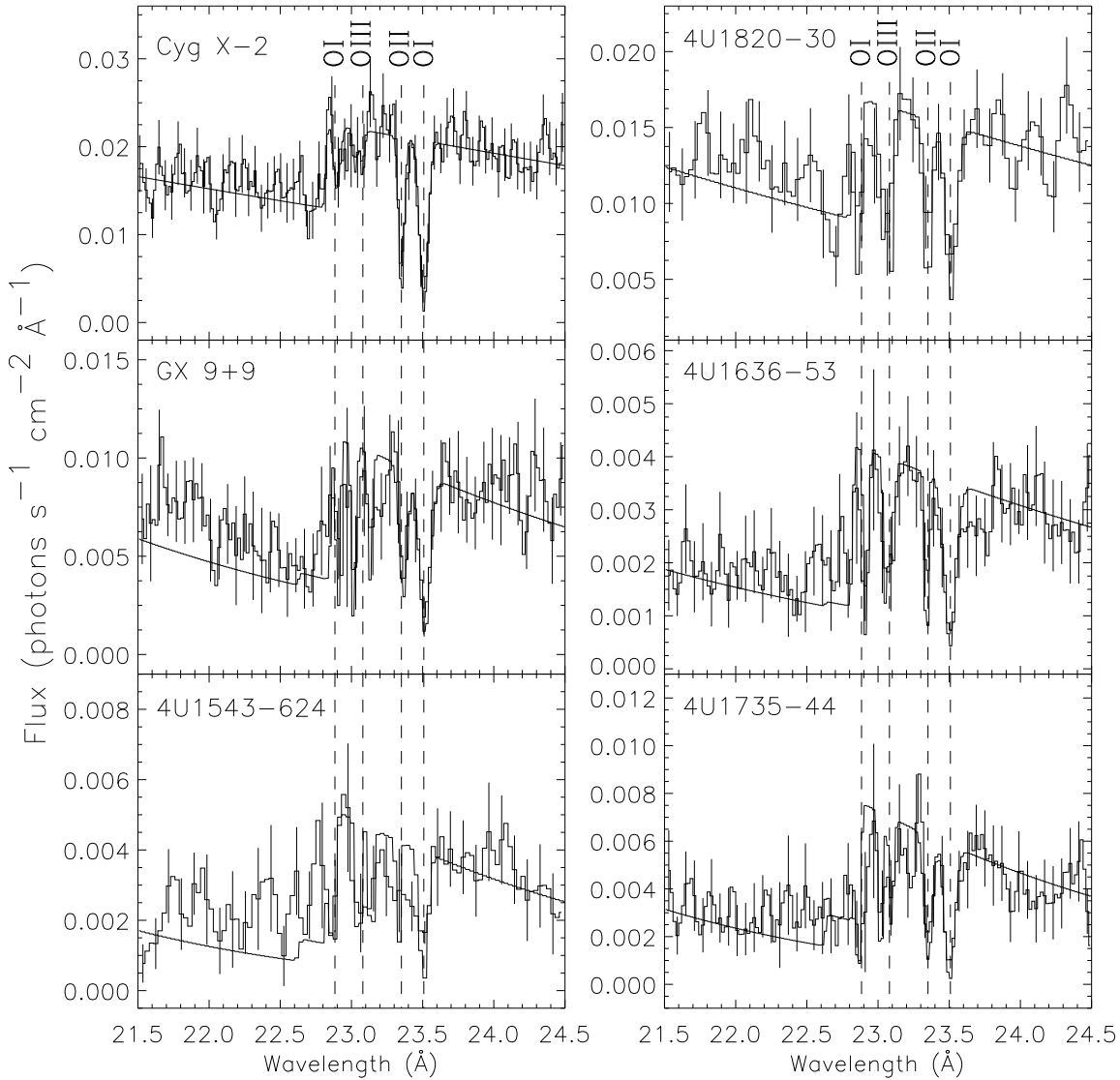} \figcaption{Flux spectra for six X-ray
binaries included in this study.  The best-fit model is also shown.
The data have been smoothed for illustrative purposes and
representative error bars are shown.  The dashed lines indicate the
positions of the identified features.}
\label{fig:4}
\end{figure*}

Finally, we turn to the two absorption features in the
23.00--23.15~\AA\/ range.  The weighted mean wavelengths of these
features are 23.112~\AA\/ and 23.034~\AA\/ (with an error of
0.004~\AA).  As shown in Table 1, the positions of the \ion{O}{3}
lines are 22.99, 23.08, and 23.14~\AA.  The resolution of the MEG is
high enough that these lines should be resolvable, but with the low
countrates, we have binned the data 1.5--3 times the instrument
resolution, which will cause the three lines to blend.  Given that we
are fitting three lines with two Gaussian absorption features, we
would expect the measured positions to lie in between the predicted
positions, which is what we found.  As a test case, we fit one of the
Cyg X-1 spectra (ObsID 3724) with three Gaussians for the \ion{O}{3}
region to determine if the lines could be reliably separated.  When
the wavelengths of the lines were fixed, significant detections of
absorption lines at those positions were found.  When the positions
were free to vary, the lines were also significantly detected and with
wavelengths consistent with the predictions of \citet{pcd+03},
although with larger errors than found when only two lines were used
to model the \ion{O}{3} absorption.  We therefore identify these
absorption features with \ion{O}{3} absorption (Figure 3).

\begin{figure}
\epsscale{1.2} \plotone{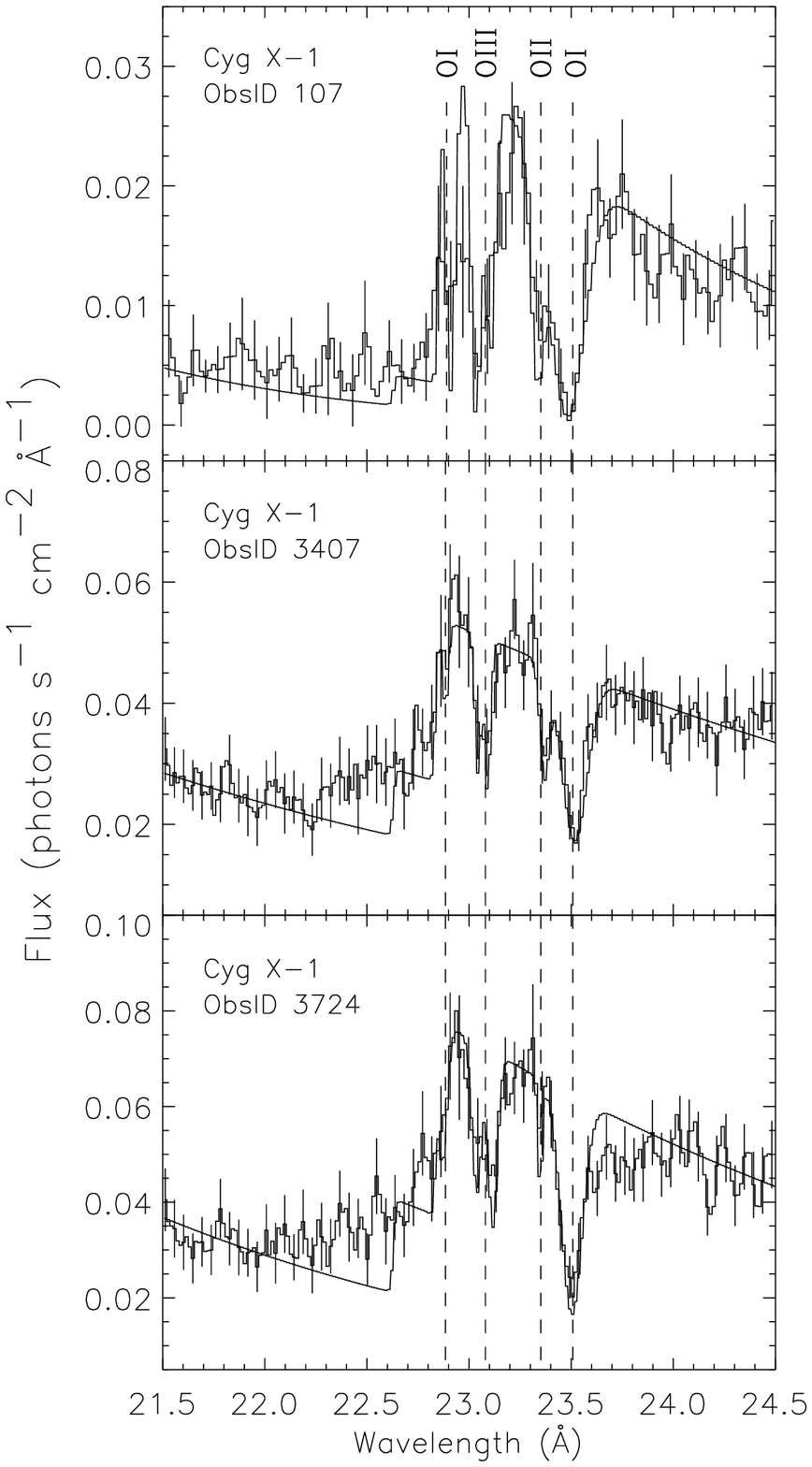} \figcaption{Flux spectra and best-fit
model for the three Cyg X-1 observations.  The data have been smoothed
for illustrative purposes and representative error bars are shown.
The dashed lines indicate the positions of the identified features.
Note the reduction in the relative strength of the \ion{O}{2} to
\ion{O}{1} absorption lines (23.35 and 23.51~\AA, respectively).  We
attribute this to possible problems in the modeling of the
time-dependent contaminant.}
\label{fig:5}
\end{figure}

We used the resonance oscillator strengths of \citet{pcd+03} to
compute the curve of growth for the ionized oxygen lines.  For
\ion{O}{3}, the total EW of the lines was used for comparison to the
data.  A curve of growth relates the EW and column density of a
particular element and ionization state.  We have not measured the
column densities for ionized oxygen, instead, we compare the EW of the
\ion{O}{2} and \ion{O}{3} lines to the column density of neutral
oxygen.  If we define the abundance $A=N_{\rm X}/N_{\rm O}$, where X
is the ionization state under consideration, we can relate the column
density of ionized oxygen to that of neutral oxygen.  In Figure 8, we
have plotted the curve of growth for \ion{O}{2} (upper panel) and
\ion{O}{3} (lower panel) for three velocity dispersions (20, 100, and
200 km~s$^{-1}$) and three abundances ($A$$=$1.0, 0.1, and 0.001).
Both \ion{O}{2} and \ion{O}{3} are in the flat part of the curve of
growth, making a determination of the abundance more difficult.  For
reasonable velocity dispersions ($\lesssim$200~km~s$^{-1}$), we can
constrain the abundance of both ionization states.  We find
\ion{O}{2}/\ion{O}{1}$\approx$0.1 and
\ion{O}{3}/\ion{O}{1}$\lesssim$0.1.

\begin{figure}
\epsscale{1.2} \plotone{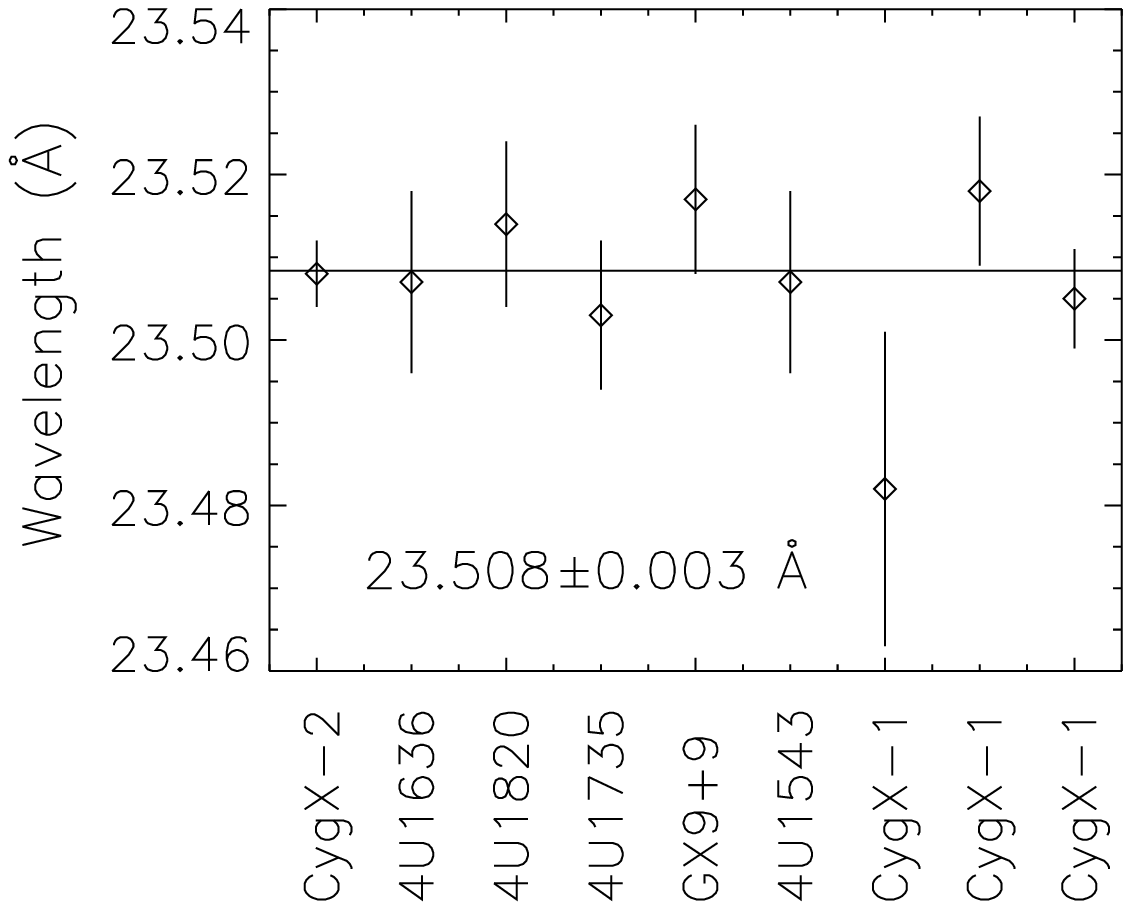} \figcaption{Best fit position of the
$1s$-$2p$ absorption line from \ion{O}{1}.  The weighted mean value of
the line position is given.  The outlier is from the Cyg X-1 ObsID 107
data.  All quoted errors are statistical errors only and do not
include the instrumental wavelength error of $\pm$0.011~\AA\/ (FWHM).}
\label{fig:6}
\end{figure}

It is interesting to note that the data points for \ion{O}{2} show a
large scatter.  The two observations with the largest deviation are
the most recent Cyg X-1 observations (ObsIDs 3407 and 3724), which are
the most affected by the time-dependent instrumental contamination.
Therefore, we suggest that the correction for the instrumental
contamination may not be completely accurate in its fine structure,
and care should be taken when drawing conclusions from the most recent
data.

\section{Discussion}
We have presented the first global study of oxygen K-shell absorption
in the ISM.  Using HETGS spectra of seven X-ray binaries, we have
developed an empirical model consisting of 2 absorption edges and 5
absorption lines.  The positions and strengths of the features were
then compared to theoretical predictions for neutral and ionized
oxygen features.  The neutral oxygen edge was resolved into the
$1s$-$2p$ and $1s$-$3p$ transitions, while two edge models were used
to describe the remaining, unresolved edge structure.  From the curve
of growth analysis, we constrain the velocity dispersion to
$\lesssim$200~km~s$^{-1}$ for neutral oxygen in the ISM.  We also
detect several absorption features which we identify with the
$1s$-$2p$ transitions of \ion{O}{2} and \ion{O}{3}.  The ISM
abundances for \ion{O}{2} and \ion{O}{3}, relative to \ion{O}{1}, are
found to be $\approx$0.1 and $\lesssim$0.1, respectively.  This is the
first measurement of the \ion{O}{2}/\ion{O}{1} and
\ion{O}{3}/\ion{O}{1} abundance of the ISM.

\begin{deluxetable*}{lcccc}
\setlength{\tabcolsep}{0.1in}
\tablecaption{Oxygen Line Parameters}
\tablehead{\colhead{Source} & \colhead{$\lambda$ (\AA)} & 
  \colhead{$\sigma$ (\AA)} & \colhead{EW (\AA)} & \colhead{$\Delta$EW (\AA)}}
\startdata
Cyg X-2 & 23.508$\pm$0.004 & 0.026$\pm$0.005 & 0.067$\pm$0.011 & \nodata \\
        & 23.352$\pm$0.005 & 0.016$\pm$0.007 & 0.041$\pm$0.011 & \nodata \\
        & 23.078$\pm$0.019 & 0.01            & 0.009$\pm$0.008 & \nodata \\
        & 23.03$\pm$0.03   & 0.01            & $<$0.017 & \nodata \\
        & 22.899$^{+0.06}_{-0.017}$ & 0.01   & 0.019$\pm$0.011 & \nodata \\ \tableline
4U~1636$-$53 & 23.507$\pm$0.011 & 0.040$^{+0.021}_{-0.014}$ & 0.09$\pm$0.03 & -0.03 \\
             & 23.340$\pm$0.008 & $<$0.02 & 0.031$\pm$0.017 & -0.003 \\
             & 23.102$^{+0.014}_{-0.06}$ & 0.01 & 0.020$\pm$0.010 & -0.005 \\ 
             & 23.056$\pm$0.013 & 0.01 & 0.027$\pm$0.010 & -0.004 \\
             & 22.915$\pm$0.013 & 0.01 & 0.034$\pm$0.010 & -0.004 \\ \tableline
4U~1820$-$30 & 23.514$\pm$0.010 & 0.034$\pm$0.015 & 0.07$\pm$0.02 & -0.02 \\
             & 23.351$\pm$0.014 & $<$0.04 & 0.04$\pm$0.02 & -0.01 \\
             & 23.086$\pm$0.017 & 0.01 & 0.033$\pm$0.013 & -0.004 \\
             & 23.03$\pm$0.06   & 0.01 & 0.023$\pm$0.013 & -0.004 \\
             & 22.867$\pm$0.015 & 0.01 & 0.036$\pm$0.016 & -0.004 \\ \tableline
4U~1735$-$44 & 23.503$\pm$0.009 & 0.033$\pm$0.015 & 0.08$\pm$0.03 & -0.02 \\
             & 23.354$\pm$0.017 & 0.027$\pm$0.014 & 0.06$\pm$0.02 & -0.03 \\
             & 23.095$\pm$0.017 & 0.01 & 0.027$\pm$0.013 & -0.016 \\
             & 23.020$\pm$0.007 & 0.01 & 0.036$\pm$0.009 & -0.010 \\  
             & 22.861$\pm$0.006 & 0.01 & 0.040$\pm$0.009 & -0.008 \\ \tableline
GX 9$+$9 & 23.517$\pm$0.009 & 0.036$\pm$0.016 & 0.08$\pm$0.03 & -0.02 \\
         & 23.367$\pm$0.017 & 0.026$\pm$0.014 & 0.050$\pm$0.019 & -0.018 \\
         & 23.133$\pm$0.010 & 0.01 & 0.024$\pm$0.010 & -0.007 \\
         & 23.017$\pm$0.011 & 0.01 & 0.034$\pm$0.010 & -0.006 \\
         & 22.906$\pm$0.018 & 0.01 & 0.031$\pm$0.011 & -0.006 \\ \tableline
4U~1543$-$624 & 23.507$\pm$0.011 & 0.025 & 0.068$\pm$0.011 & -0.005 \\
              & 23.33 & 0.01 & 0.026$\pm$0.017 & -0.009 \\
              & 23.141$\pm$0.010 & 0.01 & 0.035$\pm$0.015 & -0.012 \\
              & 23.050$\pm$0.011 & 0.01 & 0.038$\pm$0.015 & -0.010 \\
              & 22.871$\pm$0.008 & 0.01 & 0.043$\pm$0.015 & -0.009 \\ \tableline
Cyg X-1   & 23.482$\pm$0.019 & 0.086$^{+0.014}_{-0.02}$ & 0.21$\pm$0.06 & -0.06 \\
ObsID 107 & 23.333$\pm$0.015 & $<$0.05 & 0.05$^{+0.05}_{-0.03}$ & -0.001 \\
          & 23.106$\pm$0.017 & 0.01 & 0.044$\pm$0.011 & -0.009 \\
          & 23.036$\pm$0.009 & 0.01 & 0.056$\pm$0.006 & -0.002 \\
          & 22.910$\pm$0.013 & 0.01 & 0.053$\pm$0.011 & -0.008 \\ \tableline
Cyg X-1    & 23.518$\pm$0.009 & 0.061$\pm$0.017 & 0.095$\pm$0.017 & -0.018 \\
ObsID 3407 & 23.370$\pm$0.012 & 0.019$^{+0.018}_{-0.011}$ & 0.020$\pm$0.011 & -0.001 \\
           & 23.092$\pm$0.009 & 0.01 & 0.018$\pm$0.005 & -0.001 \\
           & 23.044$\pm$0.010 & 0.01 & 0.017$\pm$0.006 & -0.002 \\
           & 22.885$\pm$0.017 & 0.01 & 0.011$\pm$0.006 & -0.002 \\ \tableline
Cyg X-1    & 23.505$\pm$0.006 & 0.05 & 0.094$\pm$0.007 & -0.005 \\
ObsID 3724 & 23.349$\pm$0.010 & 0.01 & 0.009$\pm$0.005 & -0.007 \\
           & 23.118$\pm$0.010 & 0.01 & 0.030$\pm$0.006 & -0.007 \\
           & 23.039$\pm$0.018 & 0.01 & 0.024$\pm$0.006 & -0.007 \\
           & 22.870$^{+0.04}_{-0.018}$ & 0.01 & 0.022$\pm$0.014 & -0.006 \\
\vspace{-0.1in}
\enddata
\end{deluxetable*}

Our analysis relies on using the measured position of the neutral
oxygen $1s$-$2p$ line as a benchmark with which to shift the
theoretical wavelength predictions.  Previous comparisons of
theoretical predictions with laboratory data have shown good agreement
of the line positions and strengths once a small overall wavelength
offset ($\lesssim$50~m\AA) is applied \citep{mk98,gm00}.  This is at
least partially due to systematic uncertainties in the absolute
wavelength calibration of the laboratory data.  Laboratory
measurements of the $1s$-$2p$ transition of neutral oxygen range from
23.489 to 23.536~\AA\/ \citep[see][and references therein]{sls+97}
with quoted errors much less than the discrepancy among the
measurements.  In contrast, our results give a repeatable position for
the $1s$-$2p$ transition (over all our sources) due to the HETGS's
well-calibrated absolute wavelength scale.  With a wavelength accuracy
of 11~m\AA, our {\em Chandra} data is the best measurement of the
absolute positions of these features.  As an additional check, we can
compare our results with the laboratory data for singly ionized
oxygen.  \citet{kyo+02} measure the $1s$-$2p$ transition of \ion{O}{2}
at 23.35~\AA, 0.08~\AA\/ greater than predicted by \citet{pcd+03} and
consistent with our measured value.  The laboratory data supports the
need for shifts to the theoretical predictions. 

\subsection{Previous Results}
\citet{tfm+02} fit the spectrum of Cyg X-2 with three edges in the
oxygen $K$-shell region, two to represent the $^{2}P$ and $^{4}P$
edges found in atomic oxygen absorption and the other for oxygen in
molecular compounds.  The molecular edge had an edge wavelength of
23.13~\AA, similar to the position found by \citet{scc+02} for the
oxygen edge in Cyg~X-1.  This higher wavelength edge coincides with
our identified \ion{O}{3} absorption features, which could resemble an
edge feature in lower resolution spectra.  In the Cyg X-1 spectrum of
\citet{scc+02}, there is a significant count excess on the shorter
wavelength side of the edge and at the time, this excess was
attributed to either an instrumental effect or to emission lines from
the source.  Our model is more successful in fitting this structure
(see Figure 3).

Other authors \citep{pbv+01,scc+02,tfm+02} have previously identified
the \ion{O}{1} $1s$-$2p$ line and the line at 23.35~\AA\/ in the
spectra of X-ray binaries.  The 23.35~\AA\/ line has been attributed
to the $1s$-$2p$ transition in iron oxides based on the results of
\citet{wgj+97} which suggest a narrow absorption feature at
$\approx$23.3~\AA.  But in addition to the narrow absorption feature,
there is a broad edge feature at $\approx$22.9~\AA\/ \citep{wgj+97}.
In fact, given the relative cross-section between the narrow line and
the edge, the edge should be detectable before the line.  This is
opposite of the atomic oxygen cross-sections, where the narrow
$1s$-$2p$ lines have larger cross-sections than the associated edges.
This edge feature is not required by our data, which is fit well by
just the neutral, atomic oxygen edge.  Finally, the structure of
molecular edges is dependent on the composition of the dust in the
ISM, which is highly uncertain \citep{d03}.  This, coupled with a lack
of laboratory and theoretical results for molecular oxygen
cross-sections, makes predicting the position and strength of
molecular line features difficult.

We therefore attribute the 23.35~\AA\/ feature to absorption from
\ion{O}{2} since we expect ionized oxygen to be present in the ISM at
the position we determined.  In the warm, ionized phase of the ISM,
the ionization of oxygen should track that of hydrogen due to a strong
charge exchange process \citep{fs71}.  Therefore we would expect that
$N($\ion{O}{2}$)/N($\ion{O}{1}$)\approx
N($\ion{H}{2}$)/N($\ion{H}{1}$)$.  \citet{tfm+02} estimate that for
Cyg X-2, $N($\ion{H}{2}$)/N($\ion{H}{1}$)=0.32$.  \citet{r91} found
\ion{H}{2}/\ion{H}{1} of 0.2--0.6 along different lines-of-sight
through the galaxy.  Our estimate for
$N($\ion{O}{2}$)/N($\ion{O}{1}$)\approx0.1$ is less than that found
for hydrogen, but due to multiple electron ejections during
photoionization \citep{wk75}, a more accurate comparison should be
$[N($\ion{O}{2}$)+N($\ion{O}{3}$)]/N($\ion{O}{1}$)$.  In either case,
the presence of \ion{O}{2} is expected at roughly the measured value,
supporting our identification.

\subsection{Contributions from Molecules and Dust?}
Our data are well fit by our model which includes only neutral and
ionized oxygen contributions.  But the ISM should also contain oxygen
in various compounds, including carbon monoxide (CO) and dust.  In
X-ray spectra, molecular forms of oxygen should produce a $1s$-$2p$
resonance feature at a wavelength shifted from that of atomic oxygen
due to the effects of the bonds on the electron energy levels.  We
recognize that there are unidentified features in the spectra with
strengths comparable to some of the smallest EW features we identify.
It is possible that these features could be attributable to molecular
features, though any quantitative statements depend on predictions or
measurements of the strength and positions of these features, which
are unavailable at this time.

Although our data do not require such contributions, we can estimate
an upper limit to the amount allowed by our data.  If we take the
error estimate for our measured column densities to be the upper limit
on the amount of oxygen bound in molecules, and assuming the
cross-section for molecular oxygen is comparable to that for atomic,
then 10--40\% of the oxygen can be bound and still not be detected.  A
similar limit can be placed if we identify the 23.35~\AA\/ feature as
due to a molecular contribution as has been previously suggested
\citep{pbv+01,scc+02,tfm+02}.  In this case, the ratio of molecular to
atomic oxygen would be roughly the same as we suggest for
\ion{O}{2}/\ion{O}{1}, $\approx$10\%.  This assumes that the
oscillator strength for molecular oxygen, which is unknown, is similar
to that of \ion{O}{2}.  These estimates compare well with the expected
amount from other studies.  \citet{aoh+03} calculated that the
theoretical maximum amount of oxygen which can be found in dust grains
is (O/H)$_{\rm dust}\leq180$~ppm by comparing the ISM abundance of
oxygen to magnesium, silicon, and iron.  Given the measured oxygen
abundances in the ISM and the Sun, they found an average for
(O/H)$_{\rm dust}=109$~ppm, which translates to 27\% of the oxygen in
the ISM.  \citet{aoh+03} also found that the amount of oxygen bound in
grains varied between lines of sight, from 1--40\%.  Higher
signal-to-noise spectra will be required to measure the effect of dust
and molecules on the oxygen edge in X-ray spectra.

\begin{figure}
\epsscale{1.2} \plotone{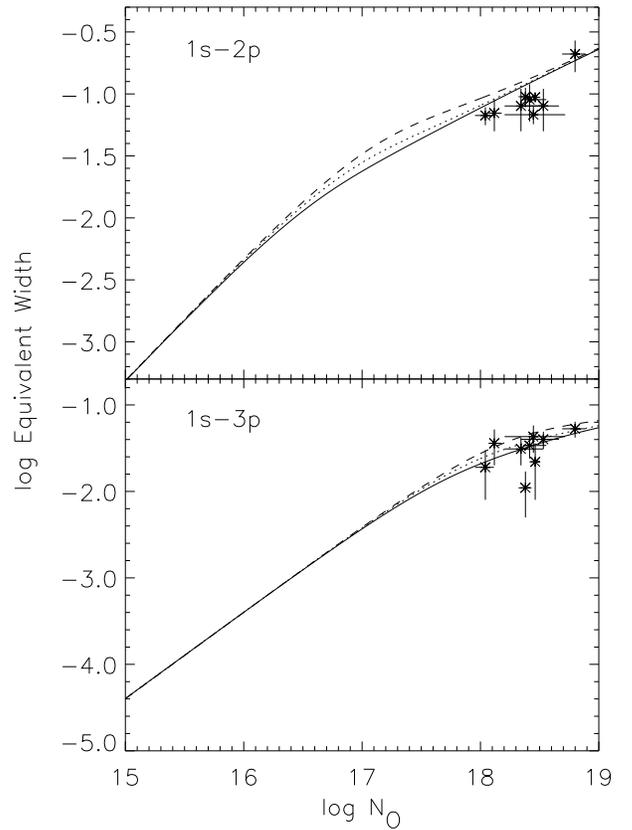} \figcaption{Curve of growth for the
$1s$-$2p$ and $1s$-$3p$ transitions of neutral oxygen.  The solid,
dashed, and dotted lines indicate the predicted EWs for velocity
dispersions of 20, 100, and 200~km~s$^{-1}$, respectively.}
\label{fig:7}
\end{figure}

\subsection{Alternatives to Interstellar Origin}
Although we have attributed the identified features to interstellar
oxygen absorption, we also consider alternative explanations for
completeness.  Pre-launch studies of the instrumental oxygen
absorption features do not show any significant features that would
account for the absorption lines we have found, other than the
23.33~\AA\/ absorption line from the OBF \citep{pbk+97}.  In addition,
the correlation of the line EWs with neutral column density would not
be seen from an instrumental feature.  Instrumental features would be
expected to give the same EW regardless of the source and would be
seen in observations of all sources, including low column density
extragalactic sources, like those used for calibration of the
contaminant contribution.  We know of no identification of these
features in lower column sources.  Therefore, we reject an
instrumental origin for our detected line features.

That being said, there does seem to be some effect on the EW
determinations of the \ion{O}{2} line from the time-dependent
contaminant absorption.  For most of our data, the contaminant effect
is small and does not impact our results for the ISM.  The most recent
observations however, which would have the largest contaminant
correction, have lower EWs than predicted from the curve of growth of
the other data points.  One possible explanation is that there is a
non-negligible \ion{O}{2} absorption line in the spectra of the
calibration source.  We note that the structure of the oxygen
contaminant absorption is primarily taken from a single observation
\citep{mtg+03}.  By not accounting for ISM \ion{O}{2}, the contaminant
correction is larger than necessary, reducing our measured EWs,
particularly as the correction grows larger.  It is hoped that our
study can help the calibration team understand ISM contributions to
the oxygen edge region in order to more accurately model the
contaminant.  Additionally, the most recent observations of Cyg X-1
also show an excess of counts in the 22.2--22.6~\AA\/ range.  Since
these are the only observations with this feature, it is again
possible that it is due to the effect of the contaminant or the
contaminant model.

\begin{figure}
\epsscale{1.2} \plotone{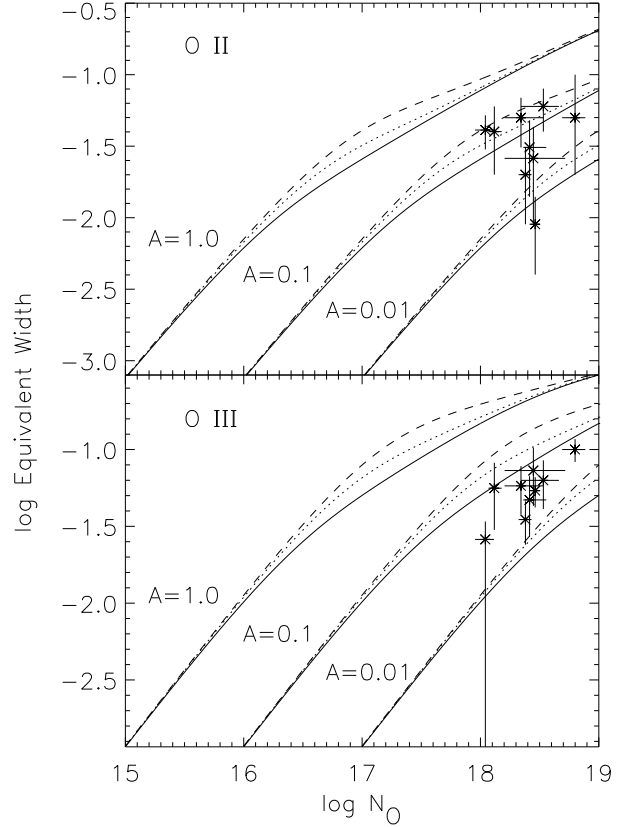} \figcaption{Curve of growth for the
\ion{O}{2} and \ion{O}{3} absorption features.  The \ion{O}{3} EW is
the total EW from the multiple absorption lines.  The solid, dashed,
and dotted lines indicate the predicted EWs for velocity dispersions
of 20, 100, and 200~km~s$^{-1}$, respectively.  Also included are
predicted EWs for various abundances ($A=N_{\rm X}/N_{\rm O}$)
relative to neutral oxygen.}
\label{fig:8}
\end{figure}

Accepting that the line features are from ionized oxygen species, we
must ask the question are these features truly from the ISM, or rather
do they originate in material local to the X-ray binaries.  The
continuum emission from the binaries provide a great source of
ionizing radiation, which could create ionization bubbles around the
systems.  The amount of ionized oxygen found in each system would
depend on the luminosity of the X-ray source, and the density and
composition of material surrounding the source.  To first order, if
the density of material is taken to be similar for all systems, the
more luminous systems will be able to ionize more oxygen and would
then be expected to have the largest measured EWs.  Of our sources,
Cyg X-1 and Cyg X-2 are the most luminous, yet Cyg X-2 has the some of
the smallest measured EWs.  This is expected if the ionized material
is interstellar, since Cyg X-2 also has the smallest neutral oxygen
column density from the edge determination.  While the \ion{O}{2} data
points do not show a significant trend with column density, most
likely due problems with modeling of the contaminant as mentioned
previously, the EWs of \ion{O}{3} are correlated with the neutral
oxygen column density.  Given that the sources included in our sample
span more than two orders of magnitude in luminosity and several
orders of magnitude in orbital separation, it is highly unlikely that
the correlation between the \ion{O}{3} EW and the neutral column
density could be produced by absorption from material local to the
binaries.  On the other hand, the correlation is exactly what is
expected if all of the absorption features are interstellar in origin.

Along with the lower ionization species found here, we would expect
the binaries to produce absorption features from more highly ionized
forms of oxygen.  Of our sources, only Cyg X-1 has shown other
absorption features \citep{scc+02,ftz03,mws+03}, and it is the only
source with multiple observations that can be used to study variations
of the column densities.  In the three observations of Cyg X-1, we
measure a change of the neutral oxygen column density of a factor of
2, which seems unreasonably high for variations in the ISM.  Cyg X-1
is a high-mass X-ray binary system that accretes from the stellar wind
of the young, massive companion.  This, combined with the detection of
other more highly ionized absorption lines, points to local material
in this system.  If the fluxes of the observations are compared, we
find that ObsID 107 has the lowest continuum flux at the oxygen edge
region and the highest neutral column density.  The fluxes are larger
in ObsIDs 3407 and 3724 with smaller columns.  We postulate that as
the flux of Cyg X-1 increases, so does the ionization of the local
material \citep[but see][for a discussion of the dependence on orbital
phase]{wcl+99,ftz03}.  In ObsIDs 3407 and 3724, the local material is
most likely highly ionized, removing any local contributions to the
neutral and low ionization features.  Measurements of the other
absorption lines may help to verify this hypothesis.

This paper represents the first in a series of studies of the
properties of the ISM using high-resolution X-ray spectra.  Future
work will include studies of lower column density sources to map out
the curve of growth of the different oxygen absorption lines.  More
data will allow us to better constrain the velocity dispersion and
ionization fraction of the oxygen in the ISM.  Additionally, we will
perform a similar analysis on the other astrophysically abundant
elements, which can be used to study abundances and depletion in the
ISM.  Studies of the ISM in X-rays provides a unique opportunity to
measure a large variety of elements and ionization states.

Some interesting questions that will be addressed in the future are
the relationship between the warm, neutral (WNM) and warm, ionized
(WIM) phases of the ISM and the source of the ionizing flux.  From an
ultraviolet observation of the Galactic halo star HD~93521,
\citet{sf93} found that the WIM coincided with the WNM.  Other studies
\citep{hs99,rtk+95} have found similar results.  It has been suggested
that the ultraviolet flux from O stars would provide enough energy to
produce the WIM \citep[see][]{hk87}.  However, the ionization fraction
of helium found by \citet{rt95} is much lower than expected for the O
star ionization models.  New observations from the {\em Far
Ultraviolet Spectroscopic Explorer} and the Wisconsin H-Alpha Mapper
will help to resolve this issue.  In addition, the high-resolution
X-ray spectra available from {\em Chandra} and {\em XMM} will provide
information on the ionization fraction of heavier elements in the ISM,
providing more data from which to understand the source of the
ionizing radiation.

\acknowledgements{We thank Claude Canizares, Carl Heiles, Frits
Paerels, and Ken Sembach for useful discussions and comments.  We
would also like to thank Tom Gorczyca for providing us with the
neutral oxygen cross-section calculation and for comments on the
manuscript.  This work was supported in part by NASA under contracts
NAS8-01129 and grant NAG5-9184.}

\end{document}